\documentstyle[12pt]{article}
\input epsf
\def\sst#1{\scriptscriptstyle{#1}}
\def\xf{x_{\!\scriptscriptstyle{f}}}
\def\thf{\theta_{\!\scriptscriptstyle{f}}}
\def\sla#1{\mbox{$#1\hspace*{-0.17cm}\scriptstyle{/}\:$}}
\newcommand{\nc}{\newcommand}
\nc{\postscript}[2] 
{\setlength{\epsfxsize}{#2\hsize}\centerline{\epsfbox{#1}}}
\nc{\bg}{B. Grzadkowski}
\nc{\non}{\nonumber}
\nc{\barx}{\bar{x}}\nc{\pbarn}{\;\hbox {pb}}\nc{\fbarn}{\;\hbox {fb}}
\nc{\hc}{\hbox {h.c.}} 
\nc{\re}{\hbox {Re}} 
\nc{\im}{\hbox {Im}}
\nc{\mev}{\hbox {MeV}} \nc{\gev}{\;\hbox {GeV}} 
\nc{\tev}{\;\hbox {TeV}}
\def\gesim{\lower0.5ex\hbox{$\:\buildrel >\over\sim\:$}} 
\def\lesim{\lower0.5ex\hbox{$\:\buildrel <\over\sim\:$}} 
\nc{\prd}[3]{{\it Phys.\ Rev.}\ {{\bf D{#1}} (#2), #3}}
\nc{\prl}[3]{{\it Phys.\ Rev.\ Lett.}\ {{\bf {#1}} (#2), #3}}
\nc{\plb}[3]{{\it Phys.\ Lett.}\ {{\bf B{#1}} (#2), #3}}
\nc{\npb}[3]{{\it Nucl.\ Phys.}\ {{\bf B{#1}} (#2), #3}}
\nc{\ptp}[3]{{\it Prog.\ Theor.\ Phys.}\ {{\bf {#1}} (#2), #3}}
\nc{\zfp}[3]{{\it Z.\ Phys.}\ {{\bf C{#1}} (#2), #3}}
\nc{\mpla}[3]{{\it Mod.\ Phys.\ Lett.}\ {{\bf A{#1}} (#2), #3}}
\nc{\rmp}[3]{{\it Rev.\ Mod.\ Phys.}\ {{\bf {#1}} (#2), #3}}
\nc{\ijmpa}[3]{{\it Int.\ J.\ Mod.\ Phys.}\
               {{\bf A{#1}} (#2), #3}}
\nc{\ttbar}{t\bar{t}}         \nc{\bbbar}{b\bar{b}}
\nc{\tanb}{\tan \beta}        \nc{\twbdec}{t\to W^+ b}
\nc{\tbwbdec}{\bar{t}\to W^- \bar{b}}
\nc{\epem}{e^+e^-}            \nc{\eett}{\epem \to \ttbar}
\nc{\sigeett}{\sigma_{e\bar{e}\to\ttbar}}
\nc{\wpwm}{W^+W^-}            \nc{\tbar}{\bar{t}}
\nc{\bbar}{\bar{b}}           \nc{\wpp}{W^+}
\nc{\mt}{m_t}    \nc{\mts}{m_t^2}   \nc{\mw}{M_W}    \nc{\mws}{M_W^2}
\nc{\mz}{M_Z}    \nc{\mzs}{M_Z^2}
\nc{\ttbardec}{\ttbar \to W^+W^-\bbbar}
\nc{\wwbb}{W^+W^-\bbbar}      \nc{\sm}{SM}
\nc{\cw}{\cos\theta_W}        \nc{\sw}{\sin\theta_W}
\nc{\sws}{\sin^2\theta_W}     \nc{\sig}{\sigma_{tot}}
\nc{\lp}{{\ell}^+}              
\nc{\lm}{{\ell}^-}
\nc{\lpm}{{\ell}^\pm}
\nc{\tb}{\stackrel{{\scriptscriptstyle (-)}}{t}}
\nc{\bb}{\stackrel{{\scriptscriptstyle (-)}}{b}}
\nc{\fb}{\stackrel{{\scriptscriptstyle (-)}}{f}}
\nc{\epsl}{\epsilon_L}        \nc{\cp}{C\!P}

\nc{\splus}{s_+}       \nc{\smin}{s_-}        \nc{\eps}{\epsilon}
\nc{\psp}{Ps_+}        \nc{\psm}{Ps_-}        \nc{\lsp}{ls_+}
\nc{\lsm}{ls_-}        \nc{\sss}{s_+s_-}      \nc{\m}{m_t}
\nc{\mq}{m_t^2}        \nc{\mr}{\frac{1}{\m}} \nc{\av}{A_{\gamma}}
\nc{\bv}{B_{\gamma}}   \nc{\az}{A_Z}          \nc{\bz}{B_Z}
\nc{\avs}{A_{\gamma}^2}\nc{\azs}{A_Z^2}       \nc{\bzs}{B_Z^2}
\nc{\dav}{\delta \! A_{\gamma}}   \nc{\dbv}{\delta \! B_{\gamma}}
\nc{\dcv}{\delta C_{\gamma}}      \nc{\ddv}{\delta \! D_{\gamma}}
\nc{\daz}{\delta \! A_Z}          \nc{\dbz}{\delta \! B_Z}
\nc{\dcz}{\delta C_Z}             \nc{\ddz}{\delta \! D_Z}
\nc{\dev}{\delta \! E_{\gamma}}   \nc{\dez}{\delta \! E_Z}
\nc{\dfv}{\delta \! F_{\gamma}}   \nc{\dfz}{\delta \! F_Z}
\nc{\rdav}{{\rm Re}(\delta \! A_{\gamma}) \:}
\nc{\rdbv}{{\rm Re}(\delta \! B_{\gamma}) \:}
\nc{\rdcv}{{\rm Re}(\delta C_{\gamma}) \:}
\nc{\rddv}{{\rm Re}(\delta \! D_{\gamma}) \:}
\nc{\rdaz}{{\rm Re}(\delta \! A_Z) \:}
\nc{\rdbz}{{\rm Re}(\delta \! B_Z) \:}
\nc{\rdcz}{{\rm Re}(\delta C_Z) \:}
\nc{\rddz}{{\rm Re}(\delta \! D_Z) \:}
\nc{\idav}{{\rm Im}(\delta \! A_{\gamma}) \:}
\nc{\idbv}{{\rm Im}(\delta \! B_{\gamma}) \:}
\nc{\idcv}{{\rm Im}(\delta C_{\gamma}) \:}
\nc{\iddv}{{\rm Im}(\delta \! D_{\gamma}) \:}
\nc{\idaz}{{\rm Im}(\delta \! A_Z) \:}
\nc{\idbz}{{\rm Im}(\delta \! B_Z) \:}
\nc{\idcz}{{\rm Im}(\delta C_Z) \:}
\nc{\iddz}{{\rm Im}(\delta \! D_Z) \:}
\nc{\cz}{(1+v_e^2)d\:\!'^2}         \nc{\ci}{v_ed\:\!'}
\nc{\ccz}{v_ed\:\!'^2}              \nc{\cci}{d\:\!'}
\nc{\dxdcos}{{d^2\sigma \over d\xf\;d\cos\thf}}
\nc{\gl}{{\mit\Gamma}_{\ell}}         \nc{\gw}{{\mit\Gamma}_W}
\nc{\gf}{{\mit\Gamma}_{\sst{f}}}      \nc{\gb}{{\mit\Gamma}_b} 
\nc{\reaf}{\re(f_2^R)}
\nc{\bet}{\beta}                \nc{\bs}{\bet^2}

\nc{\lspace}{\;\;\;\;\;\;\;\;\;\;}  \nc{\llspace}{\lspace \lspace}

\nc{\beq}{\begin{equation}}   \nc{\eeq}{\end{equation}}
\nc{\bea}{\begin{eqnarray}}   \nc{\eea}{\end{eqnarray}}
\nc{\baa}{\begin{array}}      \nc{\eaa}{\end{array}}
\nc{\bit}{\begin{itemize}}    \nc{\eit}{\end{itemize}}
\nc{\ben}{\begin{enumerate}}  \nc{\een}{\end{enumerate}}
\nc{\bce}{\begin{center}}     \nc{\ece}{\end{center}}
\begin{document}
\pagestyle{empty} \setlength{\footskip}{2.0cm}
\setlength{\oddsidemargin}{0.5cm} \setlength{\evensidemargin}{0.5cm}
\renewcommand{\thepage}{-- \arabic{page} --}
\def\mib#1{\mbox{\boldmath $#1$}}
\def\bra#1{\langle #1 |}      \def\ket#1{|#1\rangle}
\def\vev#1{\langle #1\rangle} \def\dps{\displaystyle}
% -------------------------------------------------------------------
   \def\thebibliography#1{\centerline{REFERENCES}
     \list{[\arabic{enumi}]}{\settowidth\labelwidth{[#1]}\leftmargin
     \labelwidth\advance\leftmargin\labelsep\usecounter{enumi}}
     \def\newblock{\hskip .11em plus .33em minus -.07em}\sloppy
     \clubpenalty4000\widowpenalty4000\sfcode`\.=1000\relax}\let
     \endthebibliography=\endlist
   \def\sec#1{\addtocounter{section}{1}\section*{\hspace*{-0.72cm}
     \normalsize\bf\arabic{section}.$\;$#1}\vspace*{-0.3cm}}
% -------------------------------------------------------------------
\vspace*{-1.5cm}
\begin{flushright}
$\vcenter{
\hbox{IFT-29-99}
\hbox{TOKUSHIMA 99-03} 
\hbox{hep-ph/9911505}
% \hbox{November, 1999}
}$
\end{flushright}

\vskip 0.5cm
\begin{center}
{\large\bf New Hints for Testing Anomalous Top-Quark}

\vskip 0.1cm
{\large\bf Interactions at Future Linear Colliders}
\end{center}

\vspace*{1cm}
\begin{center}
\renewcommand{\thefootnote}{\alph{footnote})}
{\sc Bohdan GRZADKOWSKI$^{\:1),\:}$}\footnote{E-mail address:
\tt bohdan.grzadkowski@fuw.edu.pl}\ and\ 
{\sc Zenr\=o HIOKI$^{\:2),\:}$}\footnote{E-mail address:
\tt hioki@ias.tokushima-u.ac.jp}
\end{center}

\vspace*{1cm}
\centerline{\sl $1)$ Institute of Theoretical Physics,\ Warsaw 
University}
\centerline{\sl Ho\.za 69, PL-00-681 Warsaw, POLAND}

\vskip 0.3cm
\centerline{\sl $2)$ Institute of Theoretical Physics,\ 
University of Tokushima}
\centerline{\sl Tokushima 770-8502, JAPAN}

\vspace*{1.4cm}
\centerline{ABSTRACT}

\vspace*{0.4cm}
\baselineskip=20pt plus 0.1pt minus 0.1pt
Angular and energy distributions for leptons and bottom quarks in the
process $\epem \to \ttbar \to {\ell}^\pm/\!\bb \cdots$ have been
calculated assuming the most general top-quark couplings. The double
distributions depend both on modification of the $\ttbar$ production
and $\tb \to \bb\!W$ decay vertices. However, the leptonic angular
distribution turned out to be totally insensitive to non-standard
parts of $Wtb$ vertex. Distributions of decay products for polarized
top quark in its rest frame have been also calculated. It has been
found that the factorization of energy and angular dependence for the
double leptonic distribution noticed earlier for the Standard Model
survives even if one allows for deviations from the V$-$A
interactions, and the SM angular leptonic distribution turned out to
be preserved by anomalous decay vertex. \\

\vspace*{0.4cm} \vfill



\newpage
%--------------------------------------------------------------------
\renewcommand{\thefootnote}{\sharp\arabic{footnote}}
%--------------------------------------------------------------------
\pagestyle{plain} \setcounter{footnote}{0}
\baselineskip=21.0pt plus 0.2pt minus 0.1pt

% 1111111111111111111111111111111111111111111111111111111111111111111
\sec{Introduction}

In spite of the fact that the top quark has been discovered already 
several years ago \cite{top} its interactions are still unknown. It
remains an open question if the top-quark couplings obey the Standard
Model (SM) scheme of the electroweak forces or there exists a
contribution from physics beyond the SM. In this letter we will try
to construct some tools which could help to answer that question at
future $\epem$ linear colliders and therefore reveal the structure of
fundamental interactions beyond the SM.  

The top quark decays immediately after being produced \cite{Bigi} and
its huge mass $m_t\simeq 174$ GeV leads to a decay width ${\mit
\Gamma}_t$ much larger than ${\mit\Lambda}_{\rm QCD}$. Therefore the
decay process is not influenced by fragmentation effects and the
decay products will provide useful information on top-quark
properties. Here we will consider distributions either of
${\ell}^\pm$ in the inclusive process $\epem \to \ttbar \to
{\ell}^\pm \cdots$ or bottom quarks from $\epem \to \ttbar \to \bb
\cdots$. We are also studying decays of polarized top quark to
$\ell^+/b +\cdots$ in its rest frame. It turns out that the analysis
of leptonic and b-quark final states are similar and could be
presented simultaneously.

% 2222222222222222222222222222222222222222222222222222222222222222222
\sec{Framework and Formalism}

We will parameterize $\ttbar$ couplings to the photon and the $Z$
boson in the following way
\begin{equation}
{\mit\Gamma}_{vt\bar{t}}^{\mu}=
\frac{g}{2}\,\bar{u}(p_t)\,\Bigl[\,\gamma^\mu \{A_v+\delta\!A_v
-(B_v+\delta\!B_v) \gamma_5 \}
+\frac{(p_t-p_{\bar{t}})^\mu}{2m_t}(\delta C_v-\delta\!D_v\gamma_5)
\,\Bigr]\,v(p_{\bar{t}}),\ \label{ff}
\label{prodff}
\end{equation}
where $g$ denotes the $SU(2)$ gauge coupling constant, $v=\gamma,Z$,
and 
\[
\av=\frac43\sw,\ \ \bv=0,\ \ 
\az=\frac1{2\cw}\Bigl(1-\frac83\sin^2\theta_W\Bigr),\ \ 
\bz=\frac1{2\cw}
\]
denote the SM contributions to the vertices. Among the above non-SM
form factors, $\delta\!A_{\gamma,Z}$, $\delta\!B_{\gamma,Z}$, $\delta
C_{\gamma,Z}$ describe $C\!P$-conserving while $\delta\!D_{\gamma,Z}$
parameterizes $C\!P$-violating interactions. Similarly, we will adopt
the following parameterization of the $Wtb$ vertex suitable for the
$t$ and $\tbar$ decays:
\begin{eqnarray}
&&\!\!{\mit\Gamma}^{\mu}_{Wtb}=-{g\over\sqrt{2}}V_{tb}\:
\bar{u}(p_b)\biggl[\,\gamma^{\mu}(f_1^L P_L +f_1^R P_R)
-{{i\sigma^{\mu\nu}k_{\nu}}\over M_W}
(f_2^L P_L +f_2^R P_R)\,\biggr]u(p_t), \non \\
\label{dectff}
&&\!\!\bar{\mit\Gamma}^{\mu}_{Wtb}=-{g\over\sqrt{2}}V_{tb}^*\:
\bar{v}(p_{\bar{t}})
\biggl[\,\gamma^{\mu}(\bar{f}_1^L P_L +\bar{f}_1^R P_R)
-{{i\sigma^{\mu\nu}k_{\nu}}\over M_W}
(\bar{f}_2^L P_L +\bar{f}_2^R P_R)\,\biggr]v(p_{\bar{b}}),~~
\label{dectbarff}
\end{eqnarray}
where $P_{L/R}=(1\mp\gamma_5)/2$, $V_{tb}$ is the $(tb)$ element of
the Kobayashi-Maskawa matrix and $k$ is the momentum of $W$. On the
other hand, it will be assumed here that interactions of leptons with
gauge bosons are properly described by the SM. Through the
calculations all fermions except the top quark will be considered as
massless. We will also neglect terms quadratic in non-standard form
factors.

Using the technique developed by Kawasaki, Shirafuji and Tsai
\cite{technique} one can derive the following formula for the
inclusive distributions of the top-quark decay product $f$ in the
process $\eett \to f + \cdots$~\cite{GH_npb}:
\beq
\frac{d^3\sigma}{d\mib{p}_{\sst{f}}/(2p_{\sst{f}}^0)}
(\epem \to f + \cdots)
=4\int d{\mit\Omega}_t
\frac{d\sigma}{d{\mit\Omega}_t}(n,0)\frac{1}{{\mit\Gamma}_t}
\frac{d^3{\mit\Gamma}_{f}}{d\mib{p}_{\sst{f}}/(2p_{\sst{f}}^0)}
(t\to f + \cdots),
\label{master}
\eeq
where ${\mit\Gamma}_t$ is the total top-quark decay width and
$d^3{\mit\Gamma}_{\sst{f}}$ is the differential decay rate for the
process considered. $d\sigma(n,0)/d{\mit\Omega}_t$ is obtained from
the angular distribution of $\ttbar$ with spins $s_+$ and $s_-$ in
$\eett$, $d\sigma(s_+,s_-)/d{\mit\Omega}_t$, by the following
replacement:
\beq
s_{+\mu} \to n^{\sst{f}}_\mu=
-\Bigl[\:g_{\mu\nu}-\frac{{p_t}_\mu{p_t}_\nu}{m_t^2}\:\Bigr]
\frac{\sum\dps{\int} d{\mit\Phi}\:\bar{B}{\mit\Lambda}_+\gamma_5
\gamma^\nu B}{\sum\dps{\int}d{\mit\Phi}\:\bar{B}{\mit\Lambda}_+ B},
\ \ \ s_{-\mu}\to 0,
\label{n-vec}
\eeq
where the matrix element for $t(s_+)\to f+\cdots$ was expressed as
$\bar{B}u_t(p_t,s_+)$, ${\mit\Lambda}_+\equiv \sla{p}_t +m_t$,
$d{\mit\Phi}$ is the relevant final-state phase-space element and
$\sum$ denotes the appropriate spin summation.

% 3333333333333333333333333333333333333333333333333333333333333333333
\sec{Distributions in $\mib{e}^+\mib{e}^-$ CM Frame}

In this section we will present results for $d^2\sigma/d\xf d\cos
\thf$ of the top-quark decay product $f$, where $f$ could be either
$\lpm$ or $\bb$, $\xf$ denotes the normalized energy of $f$ and
$\thf$ is the angle between the $e^-$ beam direction and the
direction of $f$ momentum in the $\epem$ CM frame. 

Direct calculations performed in presence of the general decay vertex
(\ref{dectff}) lead to the following result for the $n^{\sst{f}}_\mu$
vector defined in eq.(\ref{n-vec}):
\beq
n^{\sst{f}}_\mu=
\alpha^{\sst{f}}\left(g_{\mu \nu}-\frac{p_{t \mu} p_{t \nu}}{\mts}
\right)\frac{\mt}{p_t p_{\sst{f}}}p_{\sst{f}}^\nu
\label{repl}
\eeq
where for a given final state $f$, $\alpha^{\sst{f}}$ is a calculable
depolarization factor 
\beq
\alpha^{\sst{f}}=\left\{
\baa{ll}
1 &  {\rm for}\;\;f={\ell}^+  \\
  &                           \\
{\displaystyle\frac{2r-1}{2r+1}
\Bigl[\:1+\frac{8\sqrt{r}(1-r)}{(2r-1)(2r+1)}{\rm Re}(f_2^R)\:\Bigr]}
  &  {\rm for}\;\;f=b 
\eaa
\right.
\eeq
with $r \equiv (M_W/m_t)^2$.

It should be emphasized here that the above result means that there
are no corrections to the ``polarization vector'' $n^{\ell}_\mu$ for
the semileptonic top-quark decay. As it will be shown in the next
section, that has important consequences for leptonic distributions
for polarized top quark in its rest frame. On the other hand, one can
see that the corrections to $\alpha^b$ could be substantial as the
kinematical suppression factor in the leading term $2r-1(=-0.56$)
could be canceled by the appropriate contribution from the
non-standard form factor $f_2^R$.

Applying the strategy described above and adopting the general
formula for the $t\bar{t}$ distribution $d\sigma(s_+,s_-)/d{\mit
\Omega}_t$ from ref.\cite{BGH}, one obtains the following result for
the double distribution of the angle and the rescaled energy of $f$:
\beq
\frac{d^2\sigma}{d\xf d\cos\thf}
=\frac{3\pi\beta\alpha_{\mbox{\tiny EM}}^2}{2s}B_{\sst{f}}\:
\Bigl[\:{\mit\Theta}_0^{\sst{f}}(\xf)
+\cos\thf\,{\mit\Theta}_1^{\sst{f}}(\xf)
+\cos^2\thf\,{\mit\Theta}_2^{\sst{f}}(\xf) \:\Bigr],
\label{dis1}
\eeq
where $\beta$ is the top velocity, $\alpha_{\mbox{\tiny EM}}$ is the
fine structure constant and $B_{\sst{f}}$ denotes the appropriate
branching fraction. The energy dependence is specified by the
functions ${\mit\Theta}_i^{\sst{f}}(\xf)$, explicit forms of which
are shown in Appendix. They are parameterized both by production and
decay form factors.

The angular distribution\footnote{The energy distributions could be,
  of course, obtained through the integration of eq.(\ref{dis1})
  over $\cos_{\thf}$. Results for the lepton-energy spectrum
  calculated for the general form factors considered here could be
  found in ref.\cite{BGH}, while the energy spectrum for $b$ will be
  published elsewhere \cite{GH_new}.}\ 
for $f$ could be easy obtained\footnote{In the SM limit we do
  reproduce results obtained earlier by Arens and Sehgal
  \cite{as_npb}. The $C\!P$-violating contributions for the
  semileptonic decays have been compared with the results found by
  Poulose and Rindani. After correcting several misprints in their
  formula (9) in ref.\cite{PR_prd} our results agreed. In
  ref.\cite{werner} certain observables depending on $C\!P$ violation
  in the production and the decay processes have been discussed in
  the framework of the two-Higgs doublet and supersymmetric
  extensions of the SM. Implicitly, expectation values for the
  observables considered there were also sensitive to lepton and
  b-quark angular distributions.}\ 
from eq.(\ref{dis1}) by the integration over the energy of $f$:
\beq
\frac{d\sigma}{d\cos\thf}
\equiv \int_{x_-}^{x_+} \frac{d^2\sigma}{d\xf
d\cos\thf} d\xf =
\frac{3\pi\beta\alpha_{\mbox{\tiny EM}}^2}{2s}B_{\sst{f}}
\left({\mit\Omega}_0^{\sst{f}}+{\mit\Omega}_1^{\sst{f}}
\cos\thf+{\mit\Omega}_2^{\sst{f}}\cos^2\thf\right),   
\label{dis2}
\eeq
where ${\mit\Omega}_i^{\sst{f}}=\int_{x_-}^{x_+}
{\mit\Theta}_i^{\sst{f}} dx$ and $x_\pm$ define kinematical energy
range. The decay vertex is entering our double distribution,
eq.(\ref{dis1}), {\it i}) through the functions $F^{\sst{f}}(\xf)$,
$G^{\sst{f}}(\xf)$ and $H_{1,2}^{\sst{f}}(\xf)$ defined in Appendix,
and {\it ii}) through the depolarization factor $\alpha^{\sst{f}}$.
All the non-SM parts of $F^{\sst{f}}$, $G^{\sst{f}}$ and
$H_{1,2}^{\sst{f}}$ disappear upon integration over the energy $\xf$
both for ${\ell}^+$ and $b$, as it could be seen from the explicit
forms for ${\mit\Omega}_i^{\sst{f}}$ given in Appendix. As
$\alpha^{\sst{f}}=1$ for the leptonic distribution, we conclude that
the whole dependence of the lepton distribution on non-standard
structure of the top-quark decay vertex drops out through the
integration over the energy! However, one can expect substantial
modifications for the bottom-quark distribution since corrections to
$\alpha^b$ could be large.   

The fact that the angular leptonic distribution is insensitive to
corrections to the V$-$A structure of the decay vertex allows for
much more clear tests of the production vertices through measurement
of the distribution, since that way we can avoid a contamination from
non-standard structure of the decay vertex. As an illustration we
define a $C\!P$-violating asymmetry which could be constructed using
the angular distributions of $f$ and $\bar{f}$:
\beq
{\cal A}_{\sst{CP}}(\thf)= \Big[\:
{\displaystyle \frac{d\sigma^+(\thf)}{d\cos\thf}-
\frac{d\sigma^-(\pi-\thf)}{d\cos\thf}}
\:\Bigr]\Big/\Bigl[\:
{\displaystyle \frac{d\sigma^+(\thf)}{d\cos\thf}+
\frac{d\sigma^-(\pi-\thf)}{d\cos\thf}}
\:\Bigr], \label{asym}
\eeq
where $d\sigma^{+/-}$ is referring to $f$ and $\bar{f}$
distributions, respectively. Since $\thf \to \pi-\theta_{\bar{f}}$
under $C\!P$, the asymmetry defined above is a true measure of $C\!P$
violation.\footnote{Angular asymmetries have been 
  also discussed in refs.\cite{PR_prd} and \cite{PR_plb}.}\ 
It is straightforward to find that the denominator in eq.(\ref{asym})
is
\beq
\frac{d\sigma^+(\thf)}{d\cos\thf}+
\frac{d\sigma^-(\pi-\thf)}{d\cos\thf}
=2\Bigl[\:\frac{d\sigma^+(\thf)}{d\cos\thf}\:
\Bigr]^{(0)},
\eeq
where the subscript $(0)$ denotes the SM contribution to
eq.(\ref{dis2}), while the numerator becomes
\bea
&&\frac{d\sigma^+(\thf)}{d\cos\thf}-
\frac{d\sigma^-(\pi-\thf)}{d\cos\thf} =
\frac{3\pi\beta\alpha_{\mbox{\tiny EM}}^2}{2s}B_{\sst{f}}  \non\\
&&\ \ \ \ \
\times 2\left[\:\alpha^{\sst{f}}_0 
\Bigl(1-\frac{1-\bs}{2\bet}\ln\frac{1+\bet}{1-\bet}\Bigr)
\Bigl[\:(1-3\cos^2\thf)\re(F_1)-2\cos\thf\re(F_4)
\:\Bigr] \right. \non\\
&&\ \ \ \ \ \ \
-\alpha^{\sst{f}}_1(1-\beta^2)\Bigl\{\:
\Bigl(1-\frac{1}{2\bet}\ln\frac{1+\bet}{1-\bet}\Bigr)
{\rm Re}(D^{(0)}_{V\!\!A})(1-3\cos^2\thf) \non\\
&&\ \ \ \ \ \ \ \left. 
-\Bigl[\:E^{(0)}_A-(E^{(0)}_V +E^{(0)}_A)\frac{1}{2\bet}
\ln\frac{1+\bet}{1-\bet}\:\Bigr]
\cos\thf\:\Bigr\}{\rm Re}(f_2^R-\bar{f}_2^L)\:\right],
\label{asymres}
\eea
for the coefficients $F_{1,4}$ specified in Appendix, and we
expressed $\alpha^{\sst{f}}$ as $\alpha^{\sst{f}}_0 +
\alpha^{\sst{f}}_1 {\rm Re}(f_2^R)$ with
\begin{eqnarray*}
&&\alpha^{\sst{f}}_0=1,\ \ \ \ \ \ \ \ \ \ \ \
  \alpha^{\sst{f}}_1=0\ \ \ \ \ \ \ \ \ \ \ \ \ \ \ \ \,
  ({\rm for}\ f={\ell}),\\
&&\alpha^{\sst{f}}_0=\frac{2r-1}{2r+1},\ \ \ \ \;
  \alpha^{\sst{f}}_1=\frac{8\sqrt{r}(1-r)}{(1+2r)^2}\ \ \ ({\rm for}\ f=b),
\end{eqnarray*}
up to linear terms in the non-SM parameters.

\begin{figure}[h]                                                    
\postscript{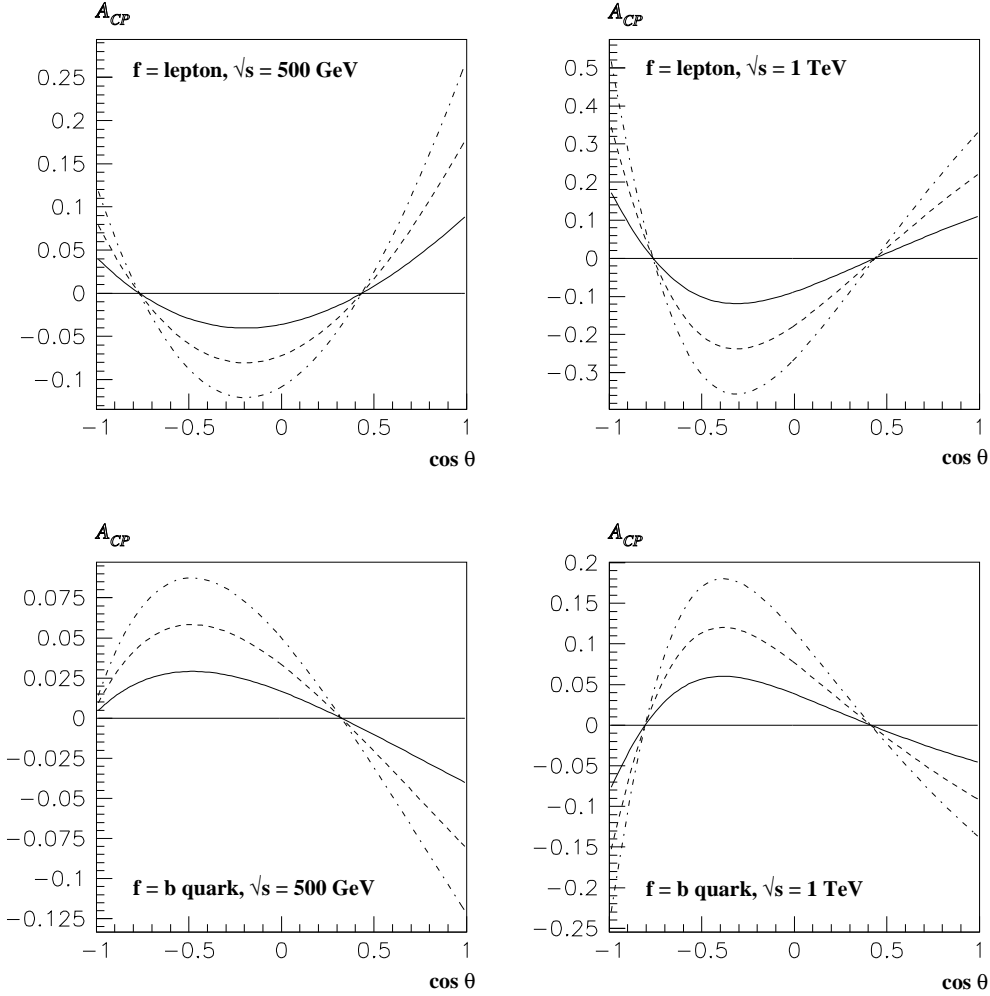}{0.95}
\vspace*{-0.5cm}    
\caption{The $C\!P$-violating asymmetry ${\cal A}_{\sst{CP}}(\thf)$
defined in eq.(\protect\ref{asym}) as a function of $\cos\thf$ for
leptonic and $b$-quark distributions for ${\rm Re}(\delta\!D_\gamma)=
{\rm Re}(\delta\!D_Z)={\rm Re}(f_2^R-\bar{f}_2^L)=$0.1 (solid line),
0.2 (dashed line), 0.3 (dash-dotted line) at $\protect\sqrt{s}=500
\gev$ and 1 TeV collider energy.}
\label{cpasym}                                                              
\end{figure}

As one could have anticipated, the asymmetry for $f={\ell}$ is
sensitive to $C\!P$ violation originating exclusively from the
production mechanism, i.e. it depends only on $F_{1,4}$ that contain
contributions from $C\!P$-violating form factors $\delta\!D_\gamma$
and $\delta\!D_Z$ while the decay vertex enters with the SM
$C\!P$-conserving coupling. For bottom quarks the effect of the
modification of the decay vertex is contained in corrections to $b$
and $\bar{b}$ depolarization factors, $\alpha^b+\alpha^{\bar{b}}=
\alpha_1^b {\rm Re}(f_2^R-\bar{f}_2^L)$ with SM $C\!P$-conserving
contribution from the production process.\footnote{One can show that
  $f_1^{L,R}=\pm\bar{f}_1^{L,R}$ and $f_2^{L,R}=\pm\bar{f}_2^{R,L}$
  where upper (lower) signs are those for $C\!P$-conserving
  (-violating) contributions \cite{cprelation}. Therefore any
  $C\!P$-violating observable defined for the top-quark decay must be
  proportional to $f_1^{L,R}-\bar{f}_1^{L,R}$ or $f_2^{L,R}-
  \bar{f}_2^{R,L}$.}\ 
As it is seen from fig.\ref{cpasym} for $\sqrt{s}=1\tev$ the
asymmetry could be quite large, e.g., reaching for the semileptonic
decays $\sim 20\%$ for ${\rm Re}(\delta\!D_\gamma)={\rm Re}(\delta\!
D_Z)=0.2$.

$C\!P$-violating form factors discussed here could be also generated
within the SM. However, it is easy to notice that first non-zero
contribution to $\delta\!D_{\gamma,Z}$ would require at least two
loops. For the top-quark decay process $C\!P$ violation could appear
at the one-loop level, however it is strongly suppressed by double
GIM mechanism \cite{GK}. Therefore we can conclude that experimental
detection of $C\!P$-violating form factors considered here would be a
clear indication for physics beyond the SM.

\newpage

% 4444444444444444444444444444444444444444444444444444444444444444444
\sec{Distributions in Top-Quark Rest Frame}
\label{restframe}

It is instructive to consider decays of a polarized top quark in its
rest frame in presence of the general decay vertex defined by
eq.(\ref{dectff}). It turns out that the leptonic angular and energy
distribution has a very similar structure to the distribution found
\cite{jez} for the pure V$-$A coupling: 
\beq
\frac{1}{\gl}\frac{d^2\gl}{dx^\star\;d\cos\theta^\star_{\ell}}=
\frac{6}{W}x^\star(1-x^\star)
\left[\:1+2\reaf\sqrt{r}\left(\frac{1}{x^\star}-\frac{3}{1+2r}\right)
\right] \frac{1+\cos\theta^\star_{\ell}}{2},
\label{disfac}
\eeq 
where $W\equiv (1-r)^2(1+2r)$, $x^\star\equiv 2E_{\ell}/\mt$ is the
normalized ${\ell}^+$ energy and $\theta^\star_{\ell}$ denotes the
angle between the top-quark spin and ${\ell}^+$ momentum.

The above formula proves that the factorization of energy and angular
dependence in the top-quark rest frame noticed by Je\.zabek and
K\"uhn for standard V$-$A top-decay vertex \cite{jez} is actually
much more general and survives even if the decay vertex given by
eq.(\ref{dectff}) is considered.\footnote{It has been found in
  ref.\cite{jezqcd} that the factorization property is approximately
  preserved by QCD corrections. The formula (\ref{disfac}) derived
  here shows that virtual one-loop QCD corrections precisely preserve
  the factorization as they just generate contributions to the form
  factor $f_2^R$.}\
One should remember that the assumptions adopted here are {\it i})
$m_b=0$ and {\it ii}) neglecting all contributions quadratic in
non-standard form factors. Under those assumptions we have proved the
factorization of energy and angular dependence.

For the angular distribution one can present results for both
${\ell}^+$ and $b$ in one formula, namely one gets
\beq
\frac{1}{\gf}\frac{d\gf}{d\cos\thf^\star}=\frac{1}{2}
[\:1+\alpha^{\sst{f}}\cos\thf^\star\:].
\label{ang}
\eeq
The coefficient $\alpha^{\sst{f}}$ which measures the amount of
information on top-quark spin direction which is being transferred to
$f$ direction is exactly the same as the depolarization factor that
appeared in eq.(\ref{repl}) in the construction of the top-quark
``polarization'' vector $n^{\sst{f}}_\mu$. We also observe that the
angular distribution (\ref{ang}) for ${\ell}^+$ is exactly the same
as for the pure V$-$A decay vertex while the one for $b$ receives
potentially large corrections from non-standard form factor $f_2^R$.
In table~\ref{alpha} we show the coefficients $\alpha^b$ calculated
for various $\re(f_2^R)$, where one can see the corrections to the SM
value of the depolarization factor reach $\sim50\%$ even for moderate
strength of the non-standard contribution to the decay vertex, such
as $\re(f_2^R)=0.1$. 
%\newpage
\renewcommand{\arraystretch}{1.4}
\begin{table}%[h]
\begin{center}
\begin{tabular}{||c||c|c|c|c|c|c|c||} 
\hline
$\re(f_2^R)$&$-$0.3&$-$0.2&$-$0.1&0.0&0.1&0.2&0.3\\ 
\hline
$\alpha^b$&$-$0.84&$-$0.70&$-$0.55&$-$0.41&$-$0.27&$-$0.12&$+$0.02 \\ 
\hline
\end{tabular}
\end{center}
\vspace*{-0.5cm}
\caption{The depolarization factor $\alpha^b$ calculated for
indicated strengths of the non-standard decay form factor
$\re(f_2^R)$.}
\label{alpha}
\end{table}

In the previous section we have noticed that the leptonic angular
distribution in $\epem$ CM was not sensitive to modifications of the
SM structure for the decay vertex. Indeed, we have observed even
though that the anomalous decay vertex was influencing the
distribution through functions $F^{\sst{f}}(\xf)$, $G^{\sst{f}}(\xf)$
and $H_{1,2}^{\sst{f}}(\xf)$ however that dependence disappeared
after integration over energy $\xf$ both for leptons and bottom
quarks. The other source of information on the decay was the
depolarization factor $\alpha^f$, which was however not modified by
non-standard interactions in the case of semileptonic decays. As we
have just seen, because of that, the angular distribution in the
top-quark rest frame was also the same as in the SM. Therefore we can
conclude that the independence of the leptonic angular distribution
in $\epem$ CM frame to corrections to the decay vertex is equivalent
to the preservation of the V$-$A form of the leptonic angular
distribution for the polarized top quark in its rest frame. The
distribution (\ref{ang}) tells us that the most likely direction of
${\ell}^+$ is the direction pointed by the top-quark spin. Since
top-quark spin is determined by the production process and the
rest-frame angular distribution is unchanged, the leptonic angular
distribution should not be sensitive to modifications of the decay
vertex. Our direct calculation confirmed that intuition.

% 5555555555555555555555555555555555555555555555555555555555555555555
\sec{Summary and Comments}

We have calculated here the angular and energy distributions both for
$\fb$ in the process $\epem \to \ttbar \to \fb \cdots$, where $f=
\ell$ or $b$ quark, assuming the most general ($C\!P$-violating {\it
and} $C\!P$-conserving) couplings for $\gamma\ttbar$, $Z \ttbar$ and
$Wtb$. The bottom-quark mass has been neglected and we have kept only
terms linear in modification of the SM vertices. We have found that
the double angular and energy distributions depend both on
modification of the $\ttbar$ production vertices as well as on
deviations from the SM at the top-quark decay vertex.

However, the angular distribution for leptons turned out to be
absolutely insensitive to variations of the standard V$-$A structure
of the $Wtb$ coupling. Therefore the distribution seems to be an
excellent tool to measure deviations from the SM in the production
process since the experimental results would not be contaminated by
unknown structure of the decay vertex. In contrast, the bottom-quark
angular distribution turned out to be influenced by non-standard
corrections to the top-quark decay vertex only through corrections to
the depolarization factor.

In order to show some $C\!P$ violating observable, we have proposed
an angular asymmetry which is sensitive to $C\!P$-violation in the
production of $\ttbar$ (for $f=\ell$) and also depends on the $C\!P$
violation parameters in top-quark decays (for $f=b$). For $\sqrt{s}=
1\tev$ colliders the asymmetry for semileptonic decays could be
substantial, e.g., reaching $\sim 20\%$ for $C\!P$-violating
production form factors of the order of $0.2$.

We have also calculated distributions of decay products for polarized
top quark in its rest frame for the same most general decay vertex.
It has been found that the factorization of energy and angular
dependence for the double distribution noticed earlier by Je\.zabek
and K\"uhn \cite{jez} survives even if one allows for deviations from
the V$-$A vertex. Since the lepton angular distribution in the rest
frame turned out to be preserved by anomalous parts of the decay
vertex, therefore our results for the angular distribution in $\epem$
CM frame could be understood qualitatively.

\vspace*{0.6cm}
% AAAAAAAAAAAAAAAAAAAAAAAAAAAAAAAAAAAAAAAAAAAAAAAAAAAAAAAAAAAAAAAAAAA
\centerline{ACKNOWLEDGMENTS}

\vspace*{0.3cm}
We would like to thank Marek Je\.zabek for helpful conversations,
Saurabh Rindani for his independent verification of the insensitivity
of the leptonic angular distribution to the $C\!P$-violating decay
vertices, and P. Poulose for correspondence on their results in
ref.\cite{PR_prd}. This work is supported in part by the State
Committee for Scientific Research (Poland) under grant 2~P03B~014~14
and by Maria Sk\l odowska-Curie Joint Fund II (Poland-USA) under
grant MEN/NSF-96-252. One of the authors (BG) is indebted to the
University of Karlsruhe, Institute of Theoretical Physics for the
warm hospitality extended to him while a part of this work was being
performed.

\vspace*{0.6cm}
% APAPAPAPAPAPAPAPAPAPAPAPAPAPAPAPAPAPAPAPAPAPAPAPAPAPAPAPAPAPAPAPAPA

\noindent \hspace*{-0.72cm}
{\bf Appendix}

Here we present explicit formulas for functions
${\mit\Theta}_i^{\sst{f}}$ describing energy dependent coefficients
for the angular and energy distributions in eq.(\ref{dis1}):
\begin{eqnarray*}
{\mit\Theta}^{\sst{f}}_0(x)\!\!&=\!\!&\Bigl[\:\frac12(3-\beta^2)D_V
   -\frac12(1-3\beta^2)D_A -(1+\beta^2){\rm Re}(G_1)  \\
&&-\alpha^{\sst{f}}[\:(1-\beta^2){\rm Re}(D_{V\!\!A}) 
     -{\rm Re}(F_1) +(2-\beta^2){\rm Re}(G_3)\:]\:\Bigr]\:
     F^{\sst{f}}(x) \\
&&+\alpha^{\sst{f}}{\rm Re}(2D_{V\!\!A}-F_1 -G_3)\:G^{\sst{f}}(x) \\
&&+\Bigl[\:D_V+D_A+2\:{\rm Re}(G_1)
+\alpha^{\sst{f}}{\rm Re}(2D_{V\!\!A}- F_1 +3G_3)\:\Bigr]\:
     H_1^{\sst{f}}(x) \\
&&-\frac12\Bigl[\:D_V+D_A+2\:{\rm Re}(G_1)
+2\alpha^{\sst{f}}\:{\rm Re}(D_{V\!\!A}+G_3)\:\Bigr]\:
     H_2^{\sst{f}}(x), \\
{\mit\Theta}^{\sst{f}}_1(x)\!\!&=\!\!&2\Bigl[\:2{\rm Re}(E_{V\!\!A})
+\alpha^{\sst{f}}[\:(1-\beta^2)E_A
-{\rm Re}(F_4-G_2)\:]\:\Bigr]F^{\sst{f}}(x) \\
&&+2\alpha^{\sst{f}}\:\Bigl[\:E_V+E_A-{\rm Re}(F_4-G_2)\:\Bigr]\:
     G^{\sst{f}}(x) \\
&&-2\:\Bigl[\:2{\rm Re}(E_{V\!\!A})+\alpha^{\sst{f}}[\:E_V+E_A
-{\rm Re}(F_4-G_2)\:]\:\Bigr]\:H_1^{\sst{f}}(x), \\
{\mit\Theta}^{\sst{f}}_2(x)\!\!&=\!\!&\Bigl[\:\frac12(3-\beta^2)
   (D_V +D_A)+(3-\beta^2){\rm Re}(G_1)\\
&&+3\alpha^{\sst{f}}[\:(1-\beta^2){\rm Re}(D_{V\!\!A})-{\rm Re}(F_1)
   +(2-\beta^2){\rm Re}(G_3)\:] \:\Bigr]\:F^{\sst{f}}(x) \\
&&+\alpha^{\sst{f}}{\rm Re}(2D_{V\!\!A}-F_1 +3G_3)\:G^{\sst{f}}(x) \\
&&-3\Bigl[\:D_V+D_A+2\:{\rm Re}(G_1)
+\alpha^{\sst{f}}{\rm Re}(2D_{V\!\!A}-F_1 +3G_3)\:\Bigr]\:
    H_1^{\sst{f}}(x) \\
&&+\frac32\Bigl[\:D_V+D_A+2\:{\rm Re}(G_1)
+2\alpha^{\sst{f}}\:{\rm Re}(D_{V\!\!A}+G_3)\:\Bigr]\:
    H_2^{\sst{f}}(x),
\end{eqnarray*}
\noindent
where
\begin{eqnarray*}
&&F^{\sst{f}}(x)\equiv\frac{1}{B_{\sst{f}}}
\int d\omega\frac1{{{\mit\Gamma}}_t}
\frac{d^2{{\mit\Gamma}}_{\sst{f}}}{dx d\omega},\ \ \
G^{\sst{f}}(x)\equiv\frac{1}{B_{\sst{f}}}\int d\omega
\Bigl[\:1-x\frac{1+\bet}{1-\omega}\Bigr]
\frac1{{{\mit\Gamma}}_t}
\frac{d^2{{\mit\Gamma}}_{\sst{f}}}{dx d\omega}, \\
&&H_1^{\sst{f}}(x)\equiv\frac{1}{B_{\sst{f}}}\frac{1-\bet}{x}
\int d\omega (1-\omega)\frac1{{{\mit\Gamma}}_t}
\frac{d^2{{\mit\Gamma}}_{\sst{f}}}{dx d\omega}, \\
&&H_2^{\sst{f}}(x)\equiv\frac{1}{B_{\sst{f}}}\Bigl(\frac{1-\bet}{x}
\Bigr)^2
\int d\omega (1-\omega)^2\frac1{{{\mit\Gamma}}_t}
\frac{d^2{{\mit\Gamma}}_{\sst{f}}}{dx d\omega},
\end{eqnarray*}
and $\omega$ is defined as $\omega\equiv (p_t -p_f)^2/m^2_t$. The
coefficients $D_V$, $D_A$, $D_{V\!\!A}$, $E_V$, $E_A$, $E_{V\!\!A}$,
$F_i$ and $G_i$ could be expressed through the production form
factors specified in eq.(\ref{prodff}), and the explicit forms of all
those relations are available in ref.\cite{BGH}. The differential
top-quark decay rates in $\epem$ CM frame, which appear in the above
definitions of $F^{\sst{f}}(x)$, $G^{\sst{f}}(x)$ and $H^{\sst{f}}_i
(x)$ are the following
$$
\frac{1}{{\mit\Gamma}_t}\frac{d^2{\mit\Gamma}_{\sst{f}}}{dx d\omega}
=\left\{
\baa{ll}
{\dps \frac{1+\beta}{\beta}\;\frac{3 B_{\ell}}{W}                         
\omega\left[\:1+2{\rm Re}(f_2^R)\sqrt{r}\left(\frac{1}{1-\omega}-      
\frac{3}{1+2r} \right)\:\right]}
 &  {\rm for}\;\;f={\ell}^+,  \\
 &                         \\
{\dps \frac{1+\beta}{2\beta(1-r)}\delta(\omega-r)}
 & 
{\rm for}\;\;f=b.
\eaa
\right.
$$
The functions appropriate for $\bar{f}$ could be obtained from the
above formulas by changing $F_{1,4}\to -F_{1,4}$, $f_2^R \to
\bar{f}_2^L$ and switching sign in front of $\cos\thf$ in
eq.(\ref{dis1}).

Integrals of ${\mit\Theta}^{\sst{f}}_i(x)$ denoted in the main text
by ${\mit\Omega}^{\sst{f}}_i$ are the following:
\begin{eqnarray*}
{\mit\Omega}^{\sst{f}}_0\!\!&=\!\!& 
D_V-(1-2\beta^2)D_A-2\:{\rm Re}(G_1)\\
&&-\alpha^{\sst{f}}[\:2(1-\beta^2){\rm Re}(D_{V\!\!A}) 
-{\rm Re}(F_1)+(3-2\beta^2){\rm Re}(G_3)\:] \\
&&+\Bigl[\:D_V+D_A+2\:{\rm Re}(G_1)
+\alpha^{\sst{f}}{\rm Re}(2D_{V\!\!A}-F_1+3G_3)\:\Bigr]
\frac{1-\beta^2}{2\beta}\ln\frac{1+\beta}{1-\beta}, \\
{\mit\Omega}^{\sst{f}}_1\!\!&=\!\!&4\:{\rm Re}(E_{V\!\!A})
+2\alpha^{\sst{f}}[\:(1-\beta^2)E_A-{\rm Re}(F_4-G_2)\:] \\
&&-\bigl\{2{\rm Re}(E_{V\!\!A})+
\alpha^{\sst{f}}[\:E_V+E_A-{\rm Re}(F_4-G_2)\:]\:\bigr\}
\frac{1-\beta^2}{\beta}\ln\frac{1+\beta}{1-\beta}, \\
{\mit\Omega}^{\sst{f}}_2\!\!&=\!\!&(3-2\beta^2)[\:D_V+D_A+
2{\rm Re}(G_1)\:]\\
&&+3\alpha^{\sst{f}}[\:2(1-\beta^2){\rm Re}(D_{V\!\!A})-{\rm Re}(F_1)
+(3-2\beta^2){\rm Re}(G_3)\:] \\
&&-3\Bigl[\:D_V+D_A+2\:{\rm Re}(G_1)
+\alpha^{\sst{f}}{\rm Re}(2D_{V\!\!A}-F_1+3G_3)\:\Bigr]
\frac{1-\beta^2}{2\beta}\ln\frac{1+\beta}{1-\beta}.
\end{eqnarray*}

\vskip 1cm
% RRRRRRRRRRRRRRRRRRRRRRRRRRRRRRRRRRRRRRRRRRRRRRRRRRRRRRRRRRRRRRRRRRR


\begin{thebibliography}{99}
%
\bibitem{top}
CDF Collaboration : F. Abe et al., \prl{73}{1994}{225};
\prd{50}{1994}{2966}; \prl{74}{1995}{2626};\\
D0 Collaboration : S. Abachi et al., \prl{74}{1995}{2632}.
%
\bibitem{Bigi}
I. Bigi and H. Krasemann, \zfp{7}{1981}{127};\\
J.H. K\"uhn, {\it Acta Phys. Austr. Suppl.} XXIV (1982), 203;\\
I. Bigi, Yu. Dokshitser, V. Khoze, J.H. K\"uhn and P. Zerwas,
\plb{181}{1986}{157}.
%
\bibitem{technique}
Y.S. Tsai, \prd{4}{1971}{2821};\\  
S. Kawasaki, T. Shirafuji and S.Y. Tsai, \ptp{49}{1973}{1656}.
%
\bibitem{GH_npb}
\bg\ and Z. Hioki, \npb{484}{1997}{17} (hep-ph/9604301);\\
see also: T. Arens and L.M. Sehgal, \prd{50}{1994}{4372}.
%
\bibitem{BGH} 
L. Brzezi\'nski, \bg\ and Z. Hioki, \ijmpa{14}{1999}{1261}
(hep-ph/9710358). 
%
\bibitem{GH_new}
\bg\ and Z. Hioki, work in progress.
%
\bibitem{as_npb} T. Arens and L.M. Sehgal, \npb{393}{1993}{46}.
%
\bibitem{PR_prd} P. Poulose and S.D. Rindani,\prd{54}{1996}{4326}
(hep-ph/9509299).
%
\bibitem{werner} W. Bernreuther and P. Overmann, \zfp{61}{1994}{599}.
%
\bibitem{PR_plb}
P. Poulose and S.D. Rindani, \plb{349}{1995}{379} (hep-ph/9410357);
\plb{383}{1996}{212} (hep-ph/9606356).
%
\bibitem{cprelation} 
W. Bernreuther, O. Nachtmann, P. Overmann, and T. Schr\"{o}der,
\npb{388}{1992}{53};\\
\bg\ and J.F. Gunion, \plb{287}{1992}{237}.
%
\bibitem{GK} \bg\ and W.-Y. Keung, \plb{319}{1993}{526}.
%
\bibitem{jez} M. Je\.zabek and J.H. K\"uhn, \npb{320}{1989}{20}.
%
\bibitem{jezqcd} M. Je\.zabek and J.H. K\"uhn, \npb{314}{1989}{1}.
%
\end{thebibliography}
\end{document}